# Entanglement Sustainability in Quantum Radar

Ahmad Salmanogli[1,2], Dincer Gokcen[1], H. Selcuk Gecim[2]

[1]Faculty of Engineering, Electrical and Electronics Engineering Department, Hacettepe University, Ankara, Turkey.

[2]Faculty of Engineering, Electrical and Electronics Engineering Department, Çankaya University, Ankara, Turkey.

**Abstract**
Quantum radar is generally defined as a detection sensor that utilizes the microwave photons like a classical radar. At the same time, it employs quantum phenomena to improve detection, identification, and resolution capabilities. However, the entanglement is so fragile, unstable, and difficult to create and to preserve for a long time. Also, more importantly, the entangled states have a tendency to leak away as a result of noise. The points mentioned above enforces that the entangled states should be carefully studied at each step of the quantum radar detection processes as follows. Firstly, the creation of the entanglement between microwave and optical photons into the tripartite system is realized. Secondly, the entangled microwave photons are intensified. Thirdly, the intensified photons are propagated into the atmosphere (attenuation medium) and reflected from a target. Finally, the backscattered photons are intensified before the detection. At each step, the parameters related to the real mediums and target material can affect the entangled states to leak away easily. In this article, the entanglement behavior of a designed quantum radar is specifically investigated.

In this study, the quantum electrodynamics theory is generally utilized to analyze the quantum radar system to define the parameters influencing the entanglement behavior. The tripartite system dynamics of equations of motions are derived using the quantum canonical conjugate method. The results of simulations indicate that the features of the tripartite system and the amplifier are designed in such a way to lead the detected photons to remain entangled with the optical modes.

**Introduction**
In recent years, the idea of using quantum phenomena in classical sensors has been used to develop the characteristics of the classical detection systems [1, 2]. It is considered to be a breakthrough utilizing quantum phenomena to improve the performance of the classical sensors such as radars [1-4], enhancing the image resolution [5-7], improving the plasmonic photodetector responsivity [8], enhancing the plasmonic system decay rate [9], and improving the Raman signals [10]. Generally, a quantum radar is recognized as a standoff sensor by which the entangled microwave photons are employed, to enhance the detection, identification, and resolution [1, 4]. Entanglement is established when two quantum particles, i.e. photons, are interacted, resulting in interrelated properties with no dependency on the distance between them [5, 11]. Quantum radars have an efficient penetration in clouds and fogs due to microwave photons which is the basic advantage of the classical radars. It has been reported that quantum radar using entangled photons dramatically enhance the resolution as compared with the non-entangled photons [1, 2]. Furthermore, the "effective visibility" of certain targets can be enhanced by using the quantum radar [1]. Also, countermeasures such as jamming can be more efficiently eliminated in case of the quantum radar that employs the entanglement [1, 2].

It has been found that all of the advantages of the quantum radar is specifically devoted to the entanglement. Nonetheless, it should be noted that the entanglement is so fragile and unstable, and also it is generally difficult to create and preserve. Additionally, entangled states have a tendency to leak away as a result of the noise [12, 13].

Naturally, the entanglement is strongly distorted due to amplifying, propagation in the atmosphere, and reflection from unknown targets, while it is employed by the quantum radar systems. Then the fragility of entangled states is dramatically increased. So, in this study, the preservation of the entanglement properties at all steps of quantum radar processing is highly emphasized. The major steps of this study are as follows: I. Creation of the entangled states by a tripartite system [14-19], II. Amplification of the entangled microwave photons [20], III. Propagation of the signal into atmosphere [1, 20], IV. Reflection from the target. The entangled states can be affected by the parameters, mentioned above, that results in leakages. Thus, in this study, the quantum radar system is designed, to minimize the leaking away of the entangled modes.

To analyze the quantum radar consisting of a tripartite system, amplifier, propagation channel, and reflection medium, the quantum electrodynamics theory is considered [21, 22], and the dynamics of quantum radar system are theoretically derived with Heisenberg-Langevin equations [23, 24]. Also, the Peres-Horodecki separability criterion is used for continuous variable system [25-28] to calculate the separability between modes.

**Description of the quantum radar system**
The schematic of the quantum radar system with some details is illustrated in scheme. 1. The process begins with the interaction of high-intensity laser with a nonlinear material to create entangled optical photon (signal and idler) [5, 11]. The idler is maintained to establish the entanglement process with the backscattered microwave photons. On the other hand, the signal excites the OC to couple the MR through the optical pressure applied on the oscillator. Oscillation of the MR resonator changes the MC capacitance by which the resonant frequency of the cavity is manipulated.

In the tripartite system, it is found that OC, MC, and MR modes couple to each other which means that the contributed modes are affected. It has been shown that, by engineering the tripartite system, the non-classical correlation can be established between the modes [14-19]. That means that the OC and MC modes can be entangled. It is the key point that we need to utilize in the quantum radar system. So, the MC output mode $C_\omega$, entangled with the optical cavity mode $a_s$, is firstly amplified through the active medium and then propagated into the atmosphere to detect the target. The active medium schematic is illustrated in scheme. 1. It contains N number of beam-splitters (BS) considered as an amplification component in the active medium. According to the quantum electrodynamics theory, the inverted thermally excited photons intensify the incoming photons at each BS [20]. So, $C_{a\omega}$ is the active medium output mode, and its entanglement should be studied with the OC modes. In fact, our design mainly concerns about preserving the entanglement at each step of the quantum radar system. The results show that the active medium properties including gain $\kappa_a$, wave vector $K_a$, and the medium length strongly affect the entanglement behavior. Second, the output modes of the active medium $C_{a\omega}$ is transmitted into the atmosphere (attenuation medium) to propagate. The attenuation medium is modeled with N discrete scattering components by which the propagated photons will be manipulated in the phase and amplitude. It is shown that the output modes of the attenuation medium $C_a$ is mainly affected by the channel length R and imaginary part of the wave vector $\kappa_{atm}$ derived by the thermally excited photons. Other critical parameters in the attenuation channel are related to the varying atmospheric conditions such as temperature and pressure [1, 20]. Consequently, the attenuation medium is one of the most important factors that can strongly manipulate the entanglement behavior in the quantum system. Third, this study also covers the reflection from the target. Similarly, quantum electrodynamics theory is employed to investigate the reflection which refers to the total amount of scattering from the target atoms. In fact, reflection is basically defined as the interaction of the electromagnetic field with the atom's quantum field. The target material is the key factor that influences the incoming photons and can strongly affect the entanglement between $C_t$ and OC modes, where $C_t$ is the reflection mode operator. After this step, the reflection photons backscatter into the atmosphere and then before the detection, they should be intensified once again. Herein, it is necessary to study the effect of the different mediums on the entanglement between the OC modes and the contributed modes. The clear relationship between the mediums and the output modes is completely illustrated in scheme. 2.

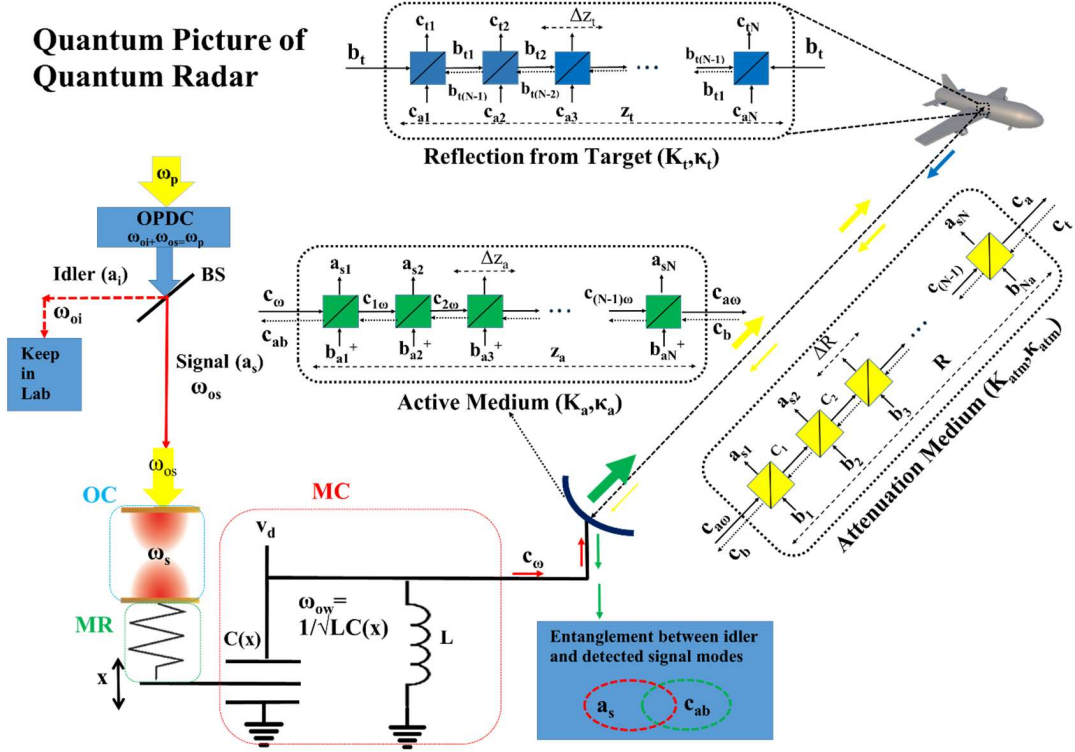

Scheme. 1 The schematic of the quantum radar; solid arrow indicates the transmitted photons, and the dashed arrows stand for backscattering the photons from the target

## Methods: Quantum theory

### A. The analysis of tripartite system with canonical conjugate methods

After a short discussion about the general operation of the system illustrated in scheme. 1, the quantum radar system is theoretically analyzed. Quantum electrodynamics theory is used to analyze the system [21, 22]. The theory starts with analyzing the dynamics equations of motion of the tripartite system, as a subsystem in quantum radar. Following the standard approach in quantum electrodynamics, the contribution of the total Lagrangian for the tripartite system is given as:

$$L_{OC} = \frac{\varepsilon_0}{2}(\dot{A}^2 - \omega_c^2 A^2)$$

$$L_{MR} = \frac{1}{2}m\dot{X}^2 - \frac{1}{2}m\omega_m^2 X^2 \quad (1)$$

$$L_{MC} = \frac{1}{2}C(x)\dot{\phi}^2 - \frac{1}{2L}\phi^2 + \frac{1}{2}C_d(v_d - \dot{\phi})^2$$

$$L_{OC-MR} = -\alpha_c A \dot{X}$$

The total Lagrangian is $L_{tot} = L_{OC} + L_{MR} + L_{MC} + L_{OC-MR}$ where $L_{OC}$, $L_{MR}$, $L_{MC}$, and $L_{OC-MR}$ are the optical cavity, microresonator, microwave cavity, and optical cavity and microresonator interaction Lagrangian, respectively. Also, $\varepsilon_0$, **A**, **X**, $\alpha_c$, and $\omega_m$, are the free space dielectric constant, vector potential, MR resonator position, interaction coefficient between OC and MR, and microresonator oscillation angular frequency, respectively. Additionally, $\omega_c$, L, **Φ**, $V_d$, C(x), $C_d$ and m optical cavity angular frequency, inductance, magnetic flux, input driving field, variable capacitor, fix capacitor between $V_d$ and LC circuit, and MR resonator mass, respectively. It is worthy to note that, the interaction Lagrangian between MC and MR is implicitly defined by the C(x) in $L_{MC}$. For the next step, one can define the conjugate variable of fields **A**, **X**, and **Φ** based on the classical definition of the conjugate variables [21], then the contributed Hamiltonian in terms of the creation and annihilation operators is expressed by:

$$H_{OC} = \hbar\omega_c \hat{a}_c^+ \hat{a}_c$$

$$H_{MR} = \frac{\hbar\omega_m}{2}(\hat{P}_x^2 + \hat{q}_x^2)$$

$$H_{MC} = \hbar\omega_\omega \hat{c}_\omega^+ \hat{c}_\omega - jV_dC_d\sqrt{\frac{\hbar\omega_\omega}{2C_t}}(\hat{c}_\omega - \hat{c}_\omega^+) \quad (2)$$

$$H_{OC-MR} = \hbar\sqrt{\frac{\alpha_c^2\omega_m}{2\varepsilon_0\omega_c m}}(\hat{a}_c^+ + \hat{a}_c)\hat{P}_x$$

$$H_{MC-MR} = C_pC_t\frac{\hbar\omega_\omega}{2}\sqrt{\frac{\hbar}{\omega_m m}}\hat{q}_x(\hat{c}_s - \hat{c}_s^+)^2$$

$$H_{oc-drive} = i\hbar E_c(\hat{a}_c^+ e^{(-j\omega_L t)} - \hat{a}_c e^{(j\omega_L t)})$$

where ($a_c^+$, $a_c$), and ($c_\omega^+$, $c_\omega$) are the creation and annihilation operators for optical and microwave cavities, respectively. Also, $\omega_L$, $P_x$, $q_x$, $E_c$, and h are the input laser angular frequency, the momentum and position normalized quadrature operator, OC cavity input driving rate, Plank's constants, respectively. Moreover, $C_p$ and $C_t$ are derived by $C_p = C'(x)/[C(x) + C_d]^2$ and $C_t = C_d + C(x_0)$, where $x_0$ indicates the equilibrium position of MR resonator. In Eq. 2, the second term in $H_{MC}$ stands for the cavity driving by the external source.

The dynamics equations of motion of the system are derived using Heisenberg-Langevin equations. Moreover, it is necessary to consider the effect of the damping rate and noise which is due to the interaction of the tripartite system with the medium. The related dynamics equation of motions of the system are given by:

$$\dot{\hat{a}}_s = -(i\Delta_c + \kappa_c)\hat{a}_c - iG_1\hat{P}_x + E_c + \sqrt{2\kappa_s}\hat{a}_{in}$$

$$\dot{\hat{c}}_\omega = -(i\Delta_\omega + \kappa_\omega)\hat{c}_\omega + i\Delta_{o\omega 1}G_2\hat{q}_x\hat{c}_\omega + E_\omega + \sqrt{2\kappa_{cs}}\hat{c}_{in} \quad (3)$$

$$\dot{\hat{q}}_x = \omega_m\hat{P}_x + G_1(\hat{a}_c^+ + \hat{a}_c)$$

$$\dot{\hat{P}}_x = -\gamma_m\hat{P}_x - \omega_m\hat{q}_x + \Delta_\omega G_2\hat{c}_\omega^+\hat{c}_\omega + \hat{b}_{in}$$

where $\kappa_c$, $\gamma_m$, and $\kappa_\omega$ are the optical cavity, microresonator, and microwave cavity damping rate, respectively. Also, $\Delta_c$, and $\Delta_\omega$ are the related detuning frequencies, and $G_1 = \sqrt{(\alpha_c^2\omega_m/2\varepsilon_0 m\omega_c)}$, $G_2 = C_pC_t\sqrt{(\hbar/m\omega_{ms})}$, and $E_\omega = V_dC_d\sqrt{(\omega_\omega/2\hbar C_t)}$. Finally, the driving field of the optical cavity $E_c$ is given as $E_c = \sqrt{(2P_c\kappa_c/\hbar\omega_L)}$ where $P_c$ is the excitation power [14-16].

A simple way to achieve a stationary and robust entanglement in continuous modes, is to select a constant point that the OC, MC, and MR are driven and worked with. Under the condition that the interaction field is so strong, it is appropriate to focus on the linearization and to calculate the quantum fluctuation around the semi-classical constant point [14, 19]. Therefore, the cavity modes are expressed as the composition of the stable and fluctuating parts as: $\mathbf{a_c} = A_s + \mathbf{\delta a_c}$, $\mathbf{c_\omega} = C_s + \mathbf{\delta c_\omega}$, $\mathbf{q_x} = X_s + \mathbf{\delta q_x}$, and $\mathbf{P_x} = P_s + \mathbf{\delta p_x}$, where the capital letter with subscript "s" denotes the stationary point, and δ indicates the fluctuation around the constant point. In the steady state condition, Eq. 3 becomes:

$$-(i\Delta_c + \kappa_c)A_s - iG_1 P_s + E_c = 0$$
$$-(i\Delta_\omega + \kappa_\omega)C_s + i\Delta_{o\omega 1} G_2 X_s C_s + E_\omega = 0 \tag{4}$$
$$\omega_m P_s + 2G_1 \text{Re}\{A_s\} = 0$$
$$-\gamma_m P_s - \omega_m X_s + \Delta_\omega G_2 |C_s|^2 = 0$$

To solve Eq. 4, it is assumed $\text{Re}\{A_s\} \gg 1$ and also $|C_s| \gg 1$. It should be noted that the fluctuation is calculated around the constant point, and the fluctuation in modes is not affected by the selection of the constant points. Moreover, it should be noted that if we choose the phase references, $A_s$ and $C_s$ can be taken real [15].

Finally, the dynamics equations of the mode fluctuations around the steady state points are given by:

$$\dot{\delta a_c} = -(i\Delta_c + \kappa_c)\hat{\delta a_c} - iG_1 \hat{\delta p_x} + \sqrt{2\kappa_s}\,\hat{\delta a_{in}}$$
$$\dot{\delta c_\omega} = -(i\Delta_\omega + \kappa_\omega)\hat{\delta c_\omega} + i\Delta_{o\omega 1} G_2 \{\hat{\delta q_x} C_s + X_s \hat{\delta c_\omega}\} + \sqrt{2\kappa_{cs}}\,\hat{\delta c_{in}} \tag{5}$$
$$\dot{\delta q_x} = \omega_m \hat{\delta p_x} + G_1(\hat{\delta a_c}^+ + \hat{\delta a_c})$$
$$\dot{\delta p_x} = -\gamma_m \hat{\delta p_x} - \omega_m \hat{\delta q_x} + \Delta_\omega G_2\{\hat{\delta c_\omega}^+ C_s + C_s^* \hat{\delta c_\omega}\} + \hat{\delta b_{in}}$$

The interaction among cavities can generate CV entanglement, i.e. quantum correlation between quadrature operator of the two intra-cavity fields and the microresonator position and momentum [9-11]. Eventually, the matrix form of the quantum electrodynamic equations of the system is introduced as:

$$\begin{bmatrix} \dot{\delta q_x} \\ \dot{\delta p_x} \\ \dot{\delta X_c} \\ \dot{\delta Y_c} \\ \dot{\delta X_\omega} \\ \dot{\delta Y_\omega} \end{bmatrix} = \underbrace{\begin{bmatrix} 0 & \omega_m & G_1\sqrt{2} & 0 & 0 & 0 \\ -\omega_m & -\gamma_m & 0 & G_m & 0 & 0 \\ 0 & 0 & -\kappa_c & \Delta_c & 0 & 0 \\ 0 & -G_1\sqrt{2} & -\Delta_c & -\kappa_c & 0 & 0 \\ G_{11} & 0 & 0 & 0 & -\kappa_{\omega 1} & \Delta_{\omega 1} \\ G_{22} & 0 & 0 & 0 & -\Delta_{\omega 1} & -\kappa_{\omega 1} \end{bmatrix}}_{A_{i,j}} \times \underbrace{\begin{bmatrix} \delta q_x \\ \delta p_x \\ \delta X_c \\ \delta Y_c \\ \delta X_\omega \\ \delta Y_\omega \end{bmatrix}}_{u(0)} + \underbrace{\begin{bmatrix} 0 \\ \delta b_{in} \\ \sqrt{2\kappa_c}\delta X_c^{in} \\ \sqrt{2\kappa_c}\delta Y_c^{in} \\ \sqrt{2\kappa_\omega}\delta X_\omega^{in} \\ \sqrt{2\kappa_\omega}\delta Y_\omega^{in} \end{bmatrix}}_{n(t)} \tag{6}$$

where $G_m = \sqrt{2}*G_2\Delta_\omega C_s$, $G_{11} = -\sqrt{2}*G_2\Delta\omega \text{Im}\{C_s\}$, $G_{22} = \sqrt{2}*G_2\Delta\omega \text{Re}\{C_s\}$, $\kappa_{\omega 1} = \kappa_\omega + G_2\Delta_\omega \text{Im}\{q_s\}$, and $\Delta_{\omega 1} = \Delta_\omega - G_2\Delta_\omega \text{Re}\{q_s\}$. Also, in Eq. 6, $\delta X_c^{in}$, $\delta Y_c^{in}$, $\delta X_\omega^{in}$, and $\delta Y_\omega^{in}$ are the quadrature operator of the related noises. The solution of Eq. 6 yields to a general form as $u(t) = \exp(A_{i,j}t)u(0) + \int(\exp(A_{i,j}s).n(t-s))ds$, where $n(s)$ is the noise column matrix. The input noises in the system obey the following correlation function [14-18].

$$<a_{in}(s)a_{in}^*(s')>=[N(\omega_c)+1]\delta(s-s'); \quad <a_{in}^*(s)a_{in}(s')>=[N(\omega_c)]\delta(s-s')$$
$$<c_{in}(s)c_{in}^*(s')>=[N(\omega_\omega)+1]\delta(s-s'); \quad <c_{in}^*(s)c_{in}(s')>=[N(\omega_\omega)]\delta(s-s')$$
$$<b_{in}(s)b_{in}^*(s')>=[N(\omega_m)+1]\delta(s-s'); \quad <b_{in}^*(s)b_{in}(s')>=[N(\omega_m)]\delta(s-s') \quad (7)$$

where $N(\omega) = [\exp(\hbar\omega/k_B T)-1]^{-1}$, $k_B$ and T are the Boltzmann's constant and operational temperature, respectively. Also, $N(\omega)$ is the equilibrium mean of the thermal photon numbers of the different modes. Solving Eq. 6 gives the cavity modes fluctuation, then one can calculate the entanglement between the modes. In a tripartite system, which is an essential part in quantum radar, the entanglement between OC and MC modes are so important. Then, in the following, we only focus on the entanglement between MC and OC modes. Also, the parameters that can destroy the non-classical correlation between the modes, are considered. To study the entanglement between two modes, one can use Simon-Peres-Horodecki criterion for continuous state separability as [25-28]:

$$\lambda_{SPH} = \det(A).\det(B) + (0.25 - |\det(C)|)^2 - tr(AJCJBJC^TJ) - 0.25 \times (\det(A) + \det(B)) \geq 0 \quad (8)$$

where J = [0, 1; -1, 0] and notation "tr" indicates the trace of the matrix. This criterion is a necessary and sufficient condition for separability of all bipartite Gaussian states system. So, if it is assumed that the contributed states in the present system are Gaussian, the entanglement of the related modes can normally be studied using Eq. 8. Moreover, A, B, and C in Eq. 8 are 2×2 correlation matrix elements [A, C; $C^T$, B] which can be presented for the case of OC-MC modes as:

$$A = \begin{bmatrix} <\delta X_c^2> - <\delta X_c>^2 & 0.5 \times <\delta X_c \delta Y_c + \delta Y_c \delta X_c> - <\delta X_c><\delta Y_c> \\ 0.5 \times <\delta X_c \delta Y_c + \delta Y_c \delta X_c> - <\delta X_c><\delta Y_c> & <\delta Y_c^2> - <\delta Y_c>^2 \end{bmatrix}$$

$$B = \begin{bmatrix} <\delta X_\omega^2> - <\delta X_\omega>^2 & 0.5 \times <\delta X_\omega \delta Y_\omega + \delta Y_\omega \delta X_\omega> - <\delta X_\omega><\delta Y_\omega> \\ 0.5 \times <\delta Y_\omega \delta X_\omega + \delta X_\omega \delta Y_\omega> - <\delta Y_\omega><\delta X_\omega> & <\delta Y_\omega^2> - <\delta Y_\omega>^2 \end{bmatrix} \quad (9)$$

$$C = \begin{bmatrix} 0.5 \times <\delta X_c \delta X_\omega + \delta X_\omega \delta X_c> - <\delta X_c><\delta X_\omega> & 0.5 \times <\delta X_c \delta Y_\omega + \delta Y_\omega \delta X_c> - <\delta X_c><\delta Y_\omega> \\ 0.5 \times <\delta Y_c \delta X_\omega + \delta X_\omega \delta Y_c> - <\delta Y_c><\delta X_\omega> & 0.5 \times <\delta Y_c \delta Y_\omega + \delta Y_\omega \delta Y_c> - <\delta Y_c><\delta Y_\omega> \end{bmatrix}$$

Therefore, it is necessary to calculate the expectation value of the quadrature modes operators (e.g. $<\delta X_c>$, $<\delta X_\omega>$, $<\delta X_c^2>$, $<\delta X_\omega^2>$, etc) to study the entanglement between modes.

So far, the tripartite system dynamics equations of motions are analytically derived using the canonical conjugate method and Heisenberg-Langevin equations. Moreover, Simon-Peres-Horodecki criterion is used to analyze the entanglement between two modes in the system. As a significant goal of this study, the tripartite system should be designed in such a way that the sustainability of the entanglement between OC-MC modes is ensured. Then, in the next step, the microwave cavity photons should be transferred into the active medium, at which the intensity of the photons is amplified to become ready for propagating into the atmosphere.

### B. The entangled Microwave photons amplifications

In this section, the details of the active medium using quantum electrodynamic theory to intensify the entrance microwave photons is explained. The main idea is to emphasis the intensifying the incoming microwave photons by preserving the nonclassical correlation behavior of the entangled photon's. The schematic of the active medium is shown in scheme. 1. The input of the active medium is microwave mode operator, $c_\omega$, and the output of the medium is $c_{a\omega}$. The purpose of the study is to investigate the effect of the active medium on entanglement between $c_{a\omega}$ and $c_\omega$. The active medium can be macroscopically modeled by a series of beam splitter (BS) connected with each other. Each BS can be operated as an amplifier by inverting the harmonic oscillator. That is associated with the thermally excited photons $b_a \rightarrow b_a^+$ where $b_a$ is the thermally excited photon operator [20]. Then the microwave signal can be intensified as propagating along the active medium by interacting with an inverted atomic population. This interaction is realized between the electromagnetic quantum and the atomic quantum fields. The coupling between the atom quantum field and the electromagnetic quantum field is expressed by the electric and magnetic dipole coupling. Using the active medium's complex coupling coefficient satisfying $|t(\omega)|^2 - |r(\omega)|^2 = 1$, the related output modes in the continuous form can be expressed as:

$$\hat{c}_{a\omega}(\omega) = e^{[iK_a + \kappa_a(\omega)]z_a} \hat{c}_\omega(\omega) + i\sqrt{2\kappa_a(\omega)} \int_0^{z_a} dz e^{[iK_a + \kappa_a(\omega)](z_a - z)} \hat{b}_a^+(\omega, z) \quad (10)$$

where $c_{a\omega}(\omega)$, $K_a$ and $\kappa_a(\omega)$ are the active medium output mode operator, the real part of the wave vector, and the imaginary part of the wave vector, respectively. It is clearly seen from Eq. 10, that the active medium intensifies the microwave cavity output mode operator (the first term), and also introduces the effect from the noise operators (the second term). The continuous form of the active medium output operators is derived in terms of input operators of MC mode. Therefore, one can study the effect of the active medium parameters on the entanglement between $c_{a\omega}$ and $a_s$. The examination of the entanglement between two modes is then followed using Eq. 8. The simulation results show that the signal intensification intensifies the either entanglement or separability. In other words, when the entangled microwave photons interact with the atom's quantum field, so the entanglement is strongly enhanced. Naturally, entanglement properties are distorted at some detuning frequencies.

### C. Propagation of the intensified photons in lossy medium

In the previous stages, firstly, design of the quantum system was theoretically analyzed by which the optical and microwave cavity modes are non-classically correlated with each other; Secondly, in order to intensify the microwave cavity signal, the amplifier was designed utilizing the quantum electrodynamics theory. Thus, the relationship was theoretically derived between the microwave cavity mode operator and the amplifier output mode. The active medium intensifies the microwave cavity mode operator by a factor being dependent on the active medium properties ($K_a$, $\kappa_a$).

Then, the intensified signal is propagated in attenuation medium, atmosphere, to detect the target. In the classical picture, there are some statistical models to simulate the atmosphere properties and study the effect of the atmosphere on the propagating signal [1]. The most critical feature of the atmosphere is its randomly changing properties. However, quantum radar has the major advantage of employing microwave photons propagating in atmosphere as compared with laser radar [1, 13, 29]. Whereas in the quantum picture, the atmosphere is considered to be a package involving the scattering centers. These centers are basically known to behave as a function of the atmospheric parameters such as temperature, pressure, and altitude. Quantum electrodynamiclly the scattering centers can be modeled with BS shown in scheme. 1. In contrast to the active medium modeling, the attenuation modeling focuses on the thermally excited photons that play an important role. It will also be shown that the latter mentioned factor drastically affects the entanglement behavior. The modeling starts with the attenuation medium scattering agent's in which for $j^{th}$ BS, the input is ($c_{a\omega j}$, $b_j$) and outputs ($c_{a\omega(j+1)}$, $a_{sj}$). Consequently, the continuous form using the attenuation medium's complex coupling coefficient satisfying $|t(\omega)|^2 + |r(\omega)|^2 = 1$, given as:

$$\hat{c}_a(\omega) = e^{[iK_{atm} - \kappa_{atm}(\omega)]R} \hat{c}_{a\omega}(\omega) + i\sqrt{2\kappa_{atm}(\omega)} \int_0^R dz e^{[iK_{atm} + \kappa_{atm}(\omega)](R-z)} \hat{b}(\omega, z) \quad (11)$$

where $c_a(\omega)$, $K_{atm}$ and $\kappa_{atm}(\omega)$ are the attenuation medium output mode operator, the real part of the wave vector, and the imaginary part of the wave vector, respectively. The effect of the attenuation channel is expressed in Eq. 11. It is readily seen that the scattering agents exponentially attenuate the input mode operator effect and also introduce the effect of the noise operator. The input operator in Eq. 11 is active medium output which was intensified in regard to Eq. 10. Following, it can be possible to examine the entanglement between two modes using Eq. 8. The simulation results show that by changing the properties of the attenuation medium, the entanglement behavior will strongly be influenced.

### D. Reflection of the microwave photons from the target

In this section, the effect of the reflection from the target is investigated on the microwave photons. In a similar way with the previous sections, the quantum electrodynamics is utilized to model the effect. The model to analyze the reflection effect is depicted in scheme. 1. The reflection from the target, in quantum picture, can be defined as the scattering from the target atoms. This process describes the interaction between the electromagnetic quantum field with the atoms quantum field. During the scattering from the target, the thermally excited photons play an important role through which the entanglement behavior can be strongly affected. To describe the process, $j^{th}$ BS is considered as a reflection agent on the target, and the related expression between the input and output operators is given in the continuous form as:

$$\hat{c}_t(\omega) = \left\{ t(\omega)e^{[iK_t \Delta z_t]} + 2\kappa_t(\omega)\sqrt{\Delta z_t} \int_0^{z_t} dz \left\{ t(\omega)e^{[iK_t - \kappa_t(\omega)]} \right\}^{(z_t-z)} \right\} \hat{c}_a(\omega) + i\sqrt{\kappa_t(\omega)\Delta z_t} e^{[iK_t - \kappa_t(\omega)]z_t} \hat{b}_t(\omega) \quad (12)$$

where $c_t(\omega)$, $K_t$ and $\kappa_t(\omega)$ are the target's scattering output mode operator, the real part of the wave vector, and the imaginary part of the wave vector, respectively. The effect of the reflection from the target is expressed in Eq. 12 in which the first term handles the direct effect of the target's imaginary part of the dielectric constant, while in the second term, the effect of the thermally excited photons is added. In other words, this equation expresses that the amplitude of the incident wave (here, it is the output of the attenuation channel) is dramatically decreased. More importantly, the phase raised due to the thermally excited photons in the second term in Eq. 12 strongly distorts the entanglement.

After the calculation of the reflection from the target, the reflected photons are scattered into the atmosphere medium. So, it is necessary to apply the atmosphere effect once again to complete the processes in quantum radar. Before the detection of the backscattered propagated photons, they should be intensified. Then, it means that an amplifier medium effect should be considered once again. It is important to emphasize that the relationship between the modes and the mediums in a typical quantum radar is illustrated in scheme. 1.

Table .1 Constant data used to simulate the system

| | |
|---|---|
| $\alpha_c$ | 0. 025~0.26 |
| $\lambda_c$ | $808*10^{-9}$ m |
| $\gamma_m$ | 120 Hz |
| m | $18~22*10^{-9}$ g |
| L | $15*10^{-12}$ H |
| $\kappa_c$ | $0.01\omega_m \sim 0.03\omega_m$ |
| $\kappa_\omega$ | $0.01\omega_m \sim 0.03\omega_m$ |
| $\omega_m$ | $2\pi*10^6$ Hz |
| $C(x_0)$ | $590*10^{-12}$ F |
| $C_d$ | $20*10^{-12}$ F |
| $P_c$ | $30*10^{-3}$ W |

**Results and Discussions**

In this section, all simulation results are explained in details. The main purpose of this study is to investigate the entanglement between modes at each step of the quantum radar and to show the effect of the related parameters addressed in scheme. 1. Some constant data used to simulate the system is tabulated in Table 1. It is worthy to note that $C_s$ and $q_s$ in Eq. 4 are assumed to be real numbers for the sake of simplicity to clearly explore the results in Fig. 1, Fig. 2, and Fig. 3.

Firstly, the tripartite cavity modes are only considered and the simulation results are illustrated in Fig. 1 and Fig. 2. In Fig. 1, the entanglement between OC-MC ($a_s$ & $c_\omega$), OC-MR ($a_s$ & $a_m$), and MR-MC ($a_m$ & $c_\omega$) are studied. This figure demonstrates a comparison among the cavity modes and it is shown that for such a design of the tripartite system, all modes become entangled around $\Delta\omega/\omega \sim 0$. Therefore, the microwave cavity photons entangled with the optical cavity photons can be transmitted into the atmosphere to detect the target.

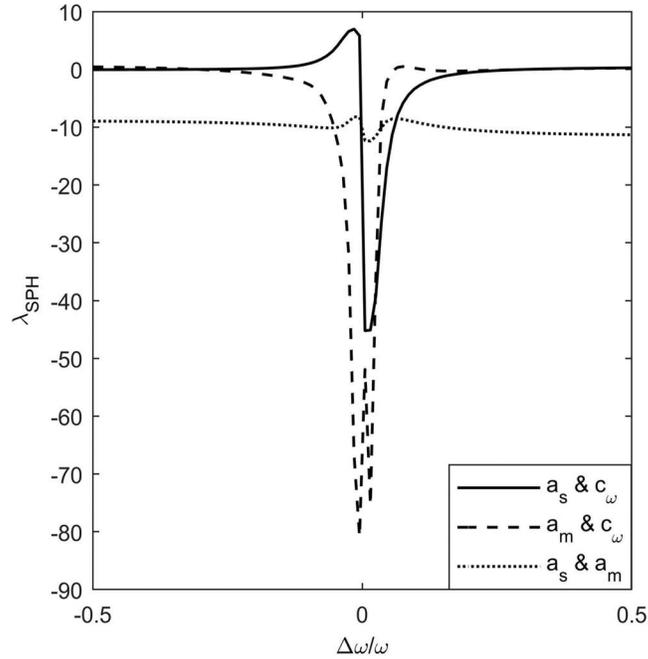

Fig. 1 Entanglement between cavities modes; T = 200 mK, $\lambda_c$ = 808 nm, m = 18 ng, $\gamma_m$ = 120 1/s.

The effect of tripartite system related parameters such as MR damping rate, incident wavelength, and temperature are investigated on the cavity modes entanglement. The effect of the temperature into the tripartite system is studied, and the results shown in Fig. 2a. These results with no surprising show that by increasing the temperature in the tripartite system, the entanglement between modes are dramatically reduced and modes become separable. The graphs related to the 200 mK and 300 mK can be seen in details in the exaggerated inset figure. The important reason that leads to the confinement of operation of the tripartite system at very low temperature is the operational frequency of the MR cavity. In fact, it is due to the relatively low frequency that MR oscillator operates with ($\omega_m = 2\pi*10^6$ Hz); from Eq. 7, it is clear that the high noise is dominant at relatively low frequency. Of course, to solve the problem, there are some studies that utilize different techniques either with engineering the frequency bandwidth [17, 18] or with replacing the MR system by the optoelectronic device that operates at high frequency [14, 19]. Therefore, to preserve the entanglement between modes, the operational temperature has to be confined below 1200 mK.

The excitation source wavelength is another important parameter that can affect the entanglement between modes in cavity. As shown in Fig. 2b, the alteration of the excitation source wavelength leads to change of the entanglement between modes. Also, in Fig. 2c, the effect of the MR damping rate is investigated on the OC and MC cavity modes, and the result reveals that the entanglement behavior is strongly affected by the change of the MR damping rate. It is contributed to the coupling factor existed between OC and MC cavities manipulated by the damping rate of the MR oscillator resonance. For better illustration of the latter mentioned point, the effect of the MR damping rate is studied on OC and MR entanglement (Fig. 2d). It is shown that by increasing the MR damping rate, the entanglement is strongly distorted due to the decrease of the coupling between MR and OC cavities.

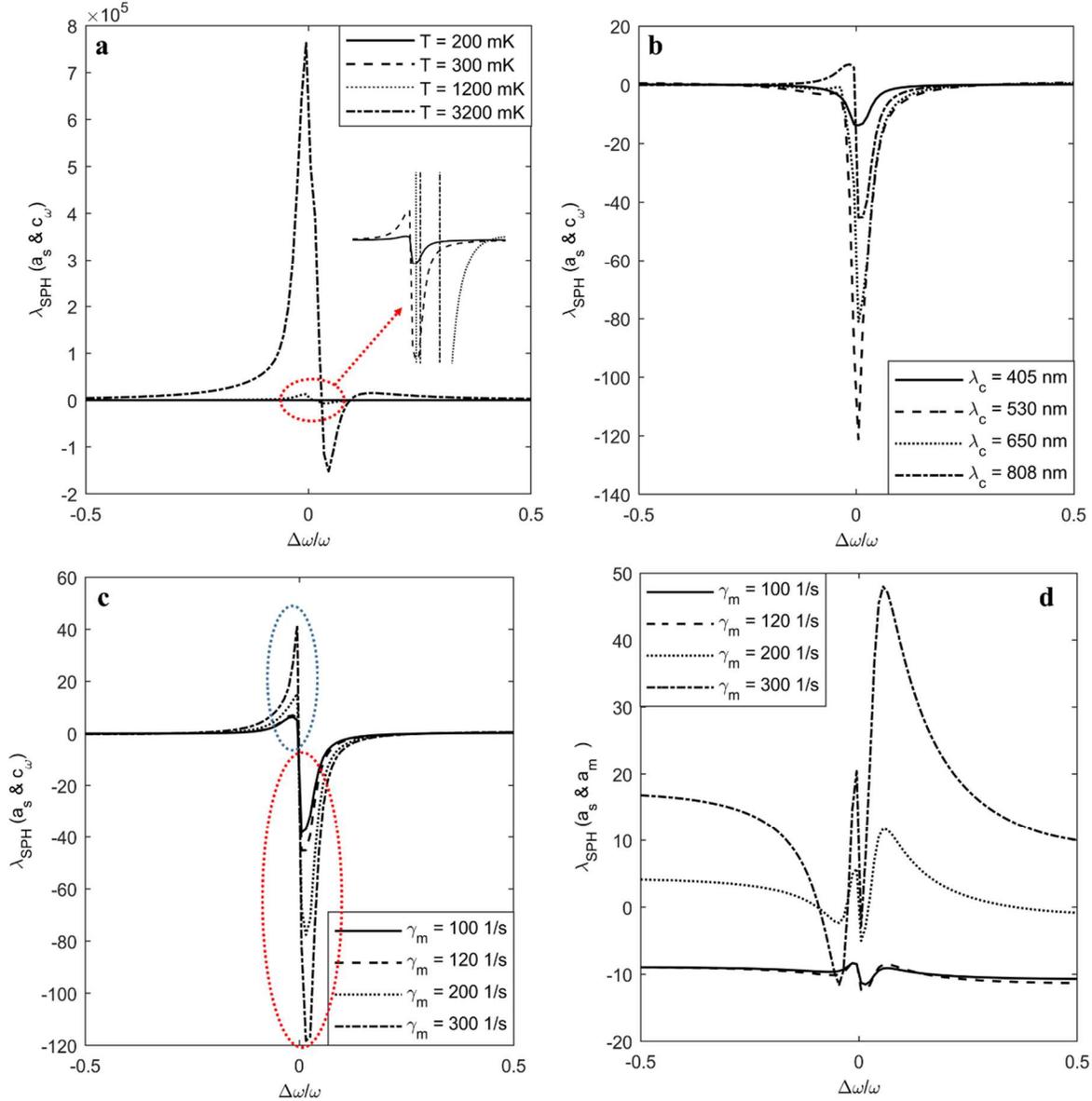

Fig. 2 Entanglement between cavities modes; $C_s$ and $q_s$ are real; (a) Temperature effect and (b) incident source wavelength effect, and (c), (d) study of the effect of the MR damping rate.

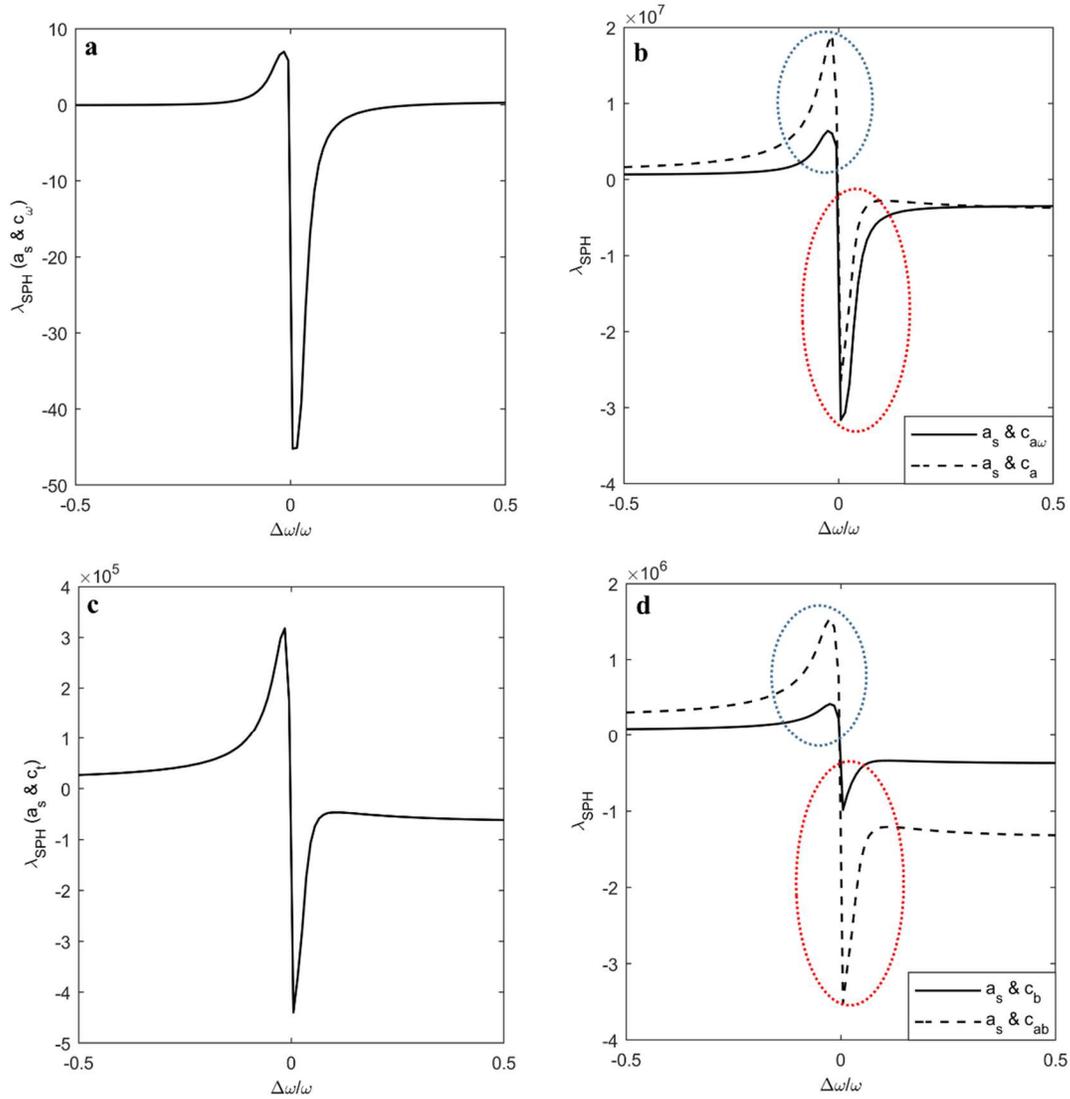

Fig. 3 Entanglement between modes at different level of quantum radar, $C_s$ and $q_s$ are real; T = 200 mK, R = 100 Km, $\lambda_c$ = 808 nm, m = 18 ng, $\gamma_m$ = 120 1/s, $\kappa_a$ = 8.7 1/m, $\kappa_{atm}$ = 5.2*10$^{-7}$ 1/m, $\kappa_t$ = 18.2 1/m.

Following the study of the tripartite system modes entanglement, and by considering the optimized conditions that the system should work with, herein, the external medium effects are evaluated for the quantum radar. In fact, the main goal is to investigate the entanglement property among different modes illustrated in scheme. 1. The design is considered for T = 200 mK, R = 100 Km, $\lambda_c$ = 808 nm, m = 18 ng, $\gamma_m$ = 120 1/s, $\kappa_a$ = 8.7 1/m, $\kappa_{atm}$ = 5.2*10$^{-7}$ 1/m, $\kappa_t$ = 18.2 1/m, depicted in Fig. 3. Based on scheme. 1, the processing starts with analyzing the cavity modes entanglement between OC and MC as illustrated in Fig. 3a. Then the microwave cavity output modes are intensified by the active medium, and the entanglement behavior of modes is studied with OC modes, as illustrated with the solid line in Fig. 3b. Interestingly, it is shown that the entanglement is slightly distorted by the active medium. Sequentially, the attenuation medium effect is considered, and the result is shown with the dashed-line in Fig. 3b. By making a comparison between solid- and dashed-lines, it is clearly seen that the separability of the modes are increased after the attenuation medium (blue dashed-circle), while the entanglement is decreased (red dashed-circle). So, it means that the attenuation medium with different conditions [29] (temperature, pressure, and so on) can strongly distort the quantum property that is critical for the sustainable entangled operation of the quantum radar. In the next step, the effect of the reflection from the target is considered. In this study, a flat target made of Aluminum is used and its refractive index is around 470 at 240 μm. The simulation result is illustrated in Fig. 3c. In comparison with the attenuation medium output, the amplitude of the signal is dramatically decreased due to the attenuation factor that the target material has. However, around $\Delta\omega/\omega \sim 0$ the resulting modes remain entangled. Of course, a target with a different shape and material can affect and distort the entanglement between modes. Following, the reflected mode from the target endures the attenuation medium effect once again, then it becomes intensified by the active medium with $\kappa_a = 0.87$ 1/m before the detection. The results of the latter mentioned effects are demonstrated in Fig. 3d. It is clearly observable that the output modes before the detection have shown entangled behavior. That is an interesting finding that we expected.

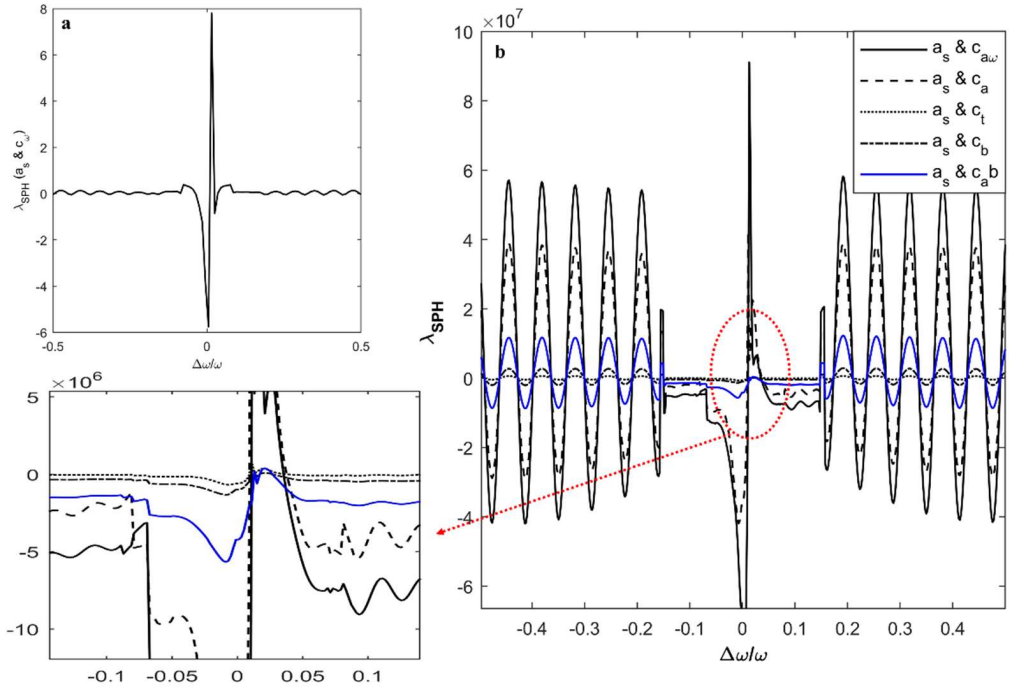

Fig. 4 Entanglement between modes at different level of quantum radar, $C_s$ and $q_s$ are complex; T = 200 mK, R = 100 Km, $\lambda_c$ = 808 nm, m = 18 ng, $\gamma_m$ = 120 1/s, $\kappa_a$ = 10.7 1/m, $\kappa_{atm}$ = 20*10$^{-7}$ 1/m, $\kappa_t$ = 18.2 1/m.

In the above simulation results, it was supposed that $C_s$ and $q_s$ in Eq. 4 are real. However, it seems to be a logical assumption with regards to choose of the phase references [15]. Nonetheless, we consider the effect of the constant's imaginary part which is an important parameter in the real system. In the following, the $C_s$ and $q_s$ are considered to be complex. Fig. 4a shows OC and MC cavities modes entanglement behavior which is especially occurred around $\Delta\omega/\omega \sim 0$ and also reveals some small oscillation at other detuning frequencies. The result demonstrates a slight difference compared with Fig. 3a. In the following, the microwave photons which is entangled with the optical cavity modes at a narrow detuning frequency should be intensified, propagated into attenuation medium, reflected from the target, backscattered into the attenuation medium, and finally intensified to be detected (Fig. 4b). From Fig. 4b, the microcavity entangled signal is initially intensified (solid black-line), indicating that either separable or entangled sections are intensified. Also, the small oscillation of microwave photons at off-detuning frequencies ($\Delta\omega/\omega \neq 0$) are strongly resonated. However, by the propagation of the output photons into the attenuation medium (dashed black-line), the entangled behavior is distorted. Then the propagated photons interact with the target and this causes to lose a lot of reflecting photons due to the absorption and scattering (dotted black-line). For better illustration, an important section of the figure is exaggerated and illustrated with the red-dashed arrow. From the exaggerated figure, the oscillation of the reflected signal with a minute amplitude and the backscattered propagated photons into the atmosphere are shown. At last, the intensified photons before the detection is depicted with blue-color in the figure. The exaggerated figure shows that the blue color signal (before detection) is not the same as the transmitted signal (solid black-line). It is because of the effect of the reflection from the target and the attenuation medium effect on the signal.

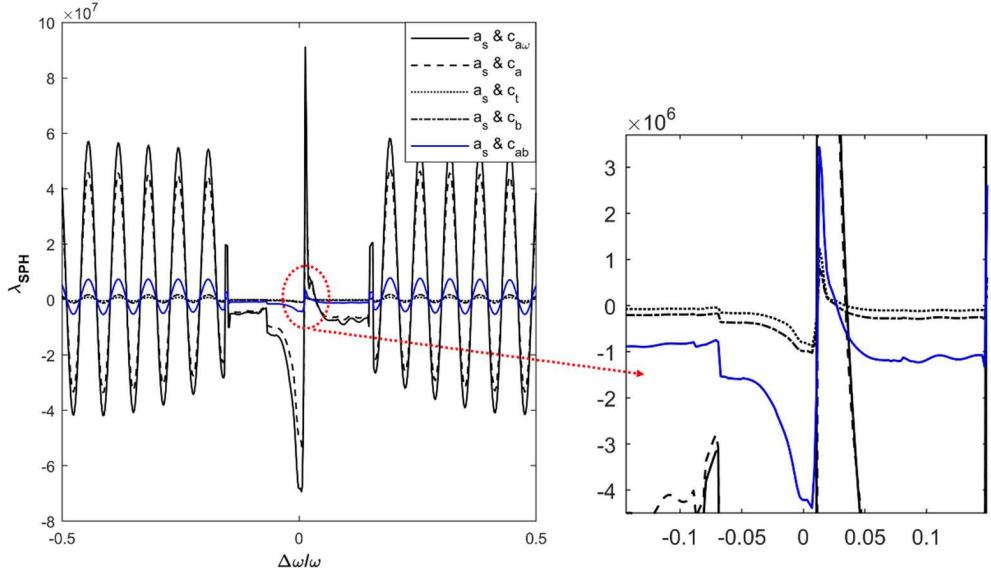

Fig. 5 Entanglement between modes at different level of quantum radar, $C_s$ and $q_s$ are complex; $T = 200$ mK, $R = 10$ Km, $\lambda_c = 808$ nm, $m = 18$ ng, $\gamma_m = 120$ 1/s, $\kappa_a = 10.7$ 1/m, $\kappa_{atm} = 100*10^{-7}$ 1/m, $\kappa_t = 18.2$ 1/m.

To compare the cases given in Fig. 4 and Fig. 5, the effect of the attenuation medium on the entanglement is studied. The distance of the attenuation medium is decreased to 10 Km. It is shown in Fig. 5 that even if $\kappa_{atm}$ is drastically increased, the blue-color graph remained undistorted and entangled. For a better illustration, an exaggerated figure is depicted in the right hand side of Fig. 5 which clearly shows the alteration of the reflected (dotted-line), propagated (dashed-dotted line) and intensified (blue solid-line) signals before the detection. Moreover, the results also show the evidence of the entanglement at off-detuning frequencies. This is due to the fact that the active medium operates with the temperatures similar to the tripartite system. In order to confirm the abovementioned points, other simulations are carried out with operational temperature at 500 mK (Fig. 6) and 1500 mK (Fig. 7). At each figure, the behavior of the entanglement at frequencies are dramatically changed as compared to Fig. 5. Those figures show a complete separable states at off-detuning frequencies. Nonetheless, the major purpose of this study is to emphasizes on-detuning frequencies $\Delta\omega/\omega \sim 0$. Fortunately, the results are appeared to be so different at on-detuning region shown in Fig. 6 and Fig. 7. One can clearly follow the effect of the different mediums on the entanglement in the exaggerated inset figures. It is also shown that by increasing temperature up to 1500 mK, the detected signal (solid blue-line) is slightly deformed due to the medium effects, however the entanglement is still maintained. Of course, it should be noted that these simulations are carried out in the very bad atmospheric conditions, with $\kappa_{atm} = 100*10^{-7}$ 1/m.

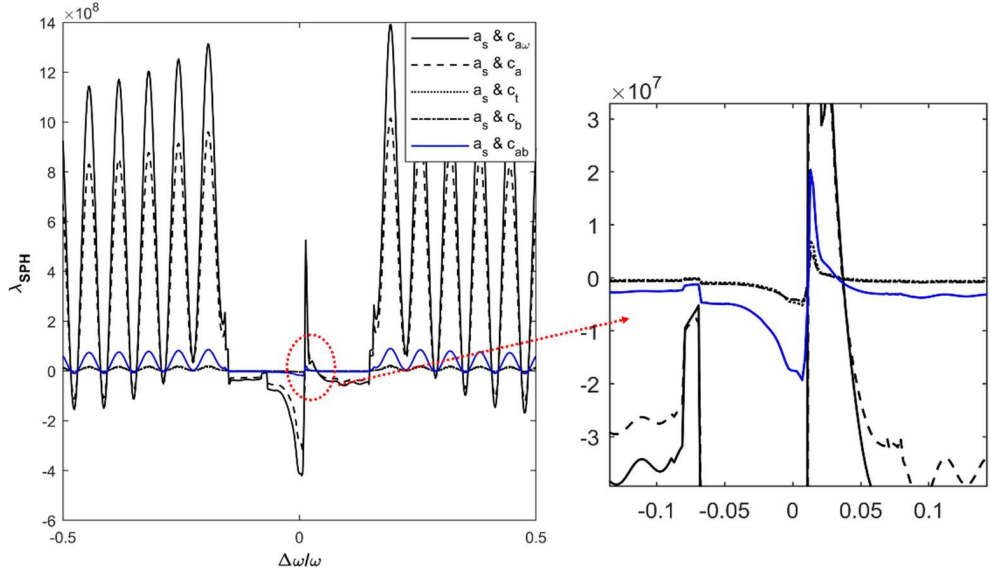

Fig. 6 Entanglement between modes at different level of quantum radar, $C_s$ and $q_s$ are complex; T = 500 mK, R = 10 Km, $\lambda_c$ = 808 nm, m = 18 ng, $\gamma_m$ = 120 1/s, $\kappa_a$ = 10.7 1/m, $\kappa_{atm}$ = 100*10$^{-7}$ 1/m, $\kappa_t$ = 18.2 1/m.

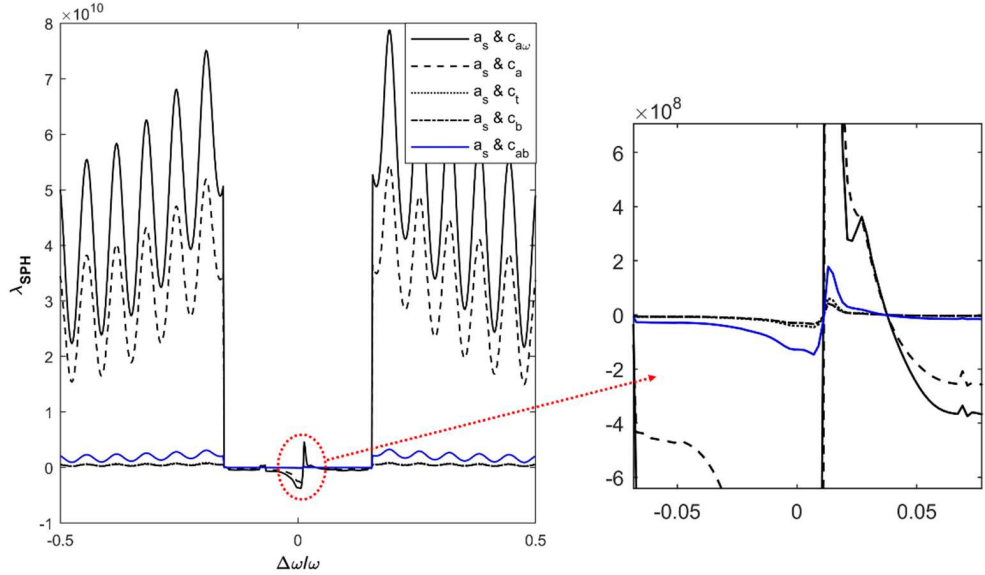

Fig. 7 Entanglement between modes at different level of quantum radar, $C_s$ and $q_s$ are complex; T = 1500 mK, R = 10 Km, $\lambda_c$ = 808 nm, m = 18 ng, $\gamma_m$ = 120 1/s, $\kappa_a$ = 10.7 1/m, $\kappa_{atm}$ = 100*10$^{-7}$ 1/m, $\kappa_t$ = 18.2 1/m.

As an important point, it should be noted that the trade-off between the estimation errors improving and modes entanglement is so critical. In other words, if one tries to reach the Heisenberg limit by enforcing to nullify the active medium gain, the entanglement between the contributed modes is distorted. This is a crucial point to be strongly considered by the quantum radar designers.

**Conclusions**
In this article, a quantum radar system was designed and examined to investigate the entanglement behavior at the generation, intensification, propagation, and reflection stages. Generally, the novelty and advantages of the quantum radar is the utilization of the entangled photons. Nevertheless, it was mentioned that the entangled states are so fragile and unstable. Thus, the entanglement behavior was completely investigated as a major purpose of the study. First of all, a tripartite cavity was designed and the related theory was derived using the canonical conjugate method. It was shown that for such a typical design, the optical and microwave cavities modes were entangled. Then, the entangled microwave photons were intensified in the active medium, propagated through an attenuation medium, and finally reflected from the target. All the different medium effects were theoretically analyzed utilizing the quantum electrodynamics theory. The results interestingly showed the sustainability of the entanglement before detection. In this study, the effect of the different parameters such as MR damping rate, temperature, atmospheric absorption coefficient, and excitation source wavelength were studied on the entanglement behavior. The results showed that if the system works at the limited temperature about 200 mK, and in the bad atmospheric condition $\kappa_{atm} = 20*10^{-7}$ 1/m, the reflected microwaves photons from the target at R = 100 Km can exhibit the entangled states around $\Delta\omega/\omega \sim 0$.